\documentclass[12pt]{article}

\usepackage{scicite}
\usepackage{graphicx}
\usepackage{times}
\usepackage{hyperref}
\usepackage{amsmath}
\usepackage{amssymb}
\usepackage{booktabs}
\usepackage{graphicx}

\newtheorem{remark}{Remark}
\newtheorem{assumption}{Assumption}
\newtheorem{theorem}{Theorem}
\newtheorem{lemma}{Lemma}

\newenvironment{sciabstract}{%
\begin{quote} \bf}
{\end{quote}}

\title{A Function-Based Approach to Model the Measurement Error in Wearable Devices }

\author
{Sneha Jadhav,$^{1\ast}$ Carmen D. Tekwe,$^{2}$ Yuanyuan Luan$^{2}$\\
\\
\normalsize{$^{1}$Department of Mathematics and Statistics, Wake Forest University,}\\
\normalsize{$^{2}$Department of Epidemiology and Biostatistics, Indiana University}\\
\\
\normalsize{$^\ast$To whom correspondence should be addressed; E-mail: jadhavs@wfu.edu.}
}

\date{}

\begin{document} 


\baselineskip24pt


\maketitle


\begin{sciabstract}
 Physical activity (PA) is an important risk factor for many health outcomes. Wearable-devices such as accelerometers are increasingly used in biomedical studies to understand the associations between PA and health outcomes. Statistical analyses involving accelerometer data are challenging due to the following three characteristics: (i) high-dimensionality, (ii) temporal dependence, and (iii) measurement error. To address these challenges we treat accelerometer- based measures of physical activity as a single function-valued covariate prone to measurement error. Specifically, in order to determine the relationship between PA and a health outcome of interest, we propose a regression model with a functional covariate that accounts for measurement error. Using regression calibration, we develop a two-step estimation method for the model parameters and establish their consistency. A test is also proposed to test the significance of the estimated model parameters. Simulation studies are conducted to compare the proposed methods with existing alternative approaches under varying scenarios. Finally, the developed methods are used to assess the relationship between PA intensity and BMI obtained from the National Health and Nutrition Examination Survey data.
\end{sciabstract}


\section*{Introduction}

It is well known that physical activity (PA) affects health. Not surprisingly, determining and characterizing relationship between PA and several health outcomes is an active research area with implications for public health. One important health outcome is overweight and obesity. There has been an alarming increase in the prevalence of overweight and obesity across the globe \cite{ng2014global}. Being major risk factors for diabetes, cardiovascular diseases along with other health conditions, overweight and obesity pose a serious public health challenge \cite{ng2014global}. Weight
gain, weight loss and weight maintenance have been linked to modifiable lifestyle behavioral factors including energy imbalance (imbalance between food intake and energy expenditure) \cite{church2018obesity}. PA plays an important role in moderating energy imbalance \cite{chung2012physical,jago2005bmi} and as a result, there is a growing interest to investigate it's role in obesity development. Accelerometers are increasingly preferred over self-report based methods to collect data on PA. These devices have the advantage over self-reported measures of PA by allowing the continuous monitoring of PA behavior over time in intervals such as epochs of 60 seconds. They monitor change in acceleration which is converted to a unit-less `counts' for a given time interval. A higher magnitude of acceleration will lead to higher counts. For further information on these devices refer to John \cite{john2012actigraph}. In short, accelerometers capture intensity of activity at several time points rendering a detailed activity profile of the individuals. In this work, we propose a method that will allow us to study the relationship between body mass index (BMI) and the accelerometer-based PA activity measures from the National Health and Nutrition Examination Survey (NHANES) data set. The novelty of this method is that it accommodates and exploits the special features of accelerometer data. This method can be used in studies involving other health health outcomes or in studies containing data from other types of wearable devices.

The NHANES \url{https://www.cdc.gov/nchs/nhanes/about\_nhanes.htm}  is an ongoing program that aims to study health and nutrition. In 2003, the tracking of PA through the use of PA monitors (PAM) was added to the NHANES. More information can be found at \url{https://wwwn.cdc.gov/nchs/nhanes/2003-2004/PAXRAW_C.htm}. Specifically, the device used was ActiGraph AM-7164,  manufactured by ActiGraph of Ft. Walton Beach, FL. It is a uniaxial accelerometer \cite{kozey2010comparison} that records the intensity of movement along with the step count at one minute intervals. The participants were asked to wear the device for 7 consecutive days during waking hours with the exception of swimming and bathing. However, non compliance resulted in several subjects having less or more wear-time than expected \cite{leroux2019organizing}. 
To reduce the noise associated with the frequency of data collection, we summarized the activity information obtained in epochs of 60 seconds to hourly level data. The hourly PA intensity of a randomly selected individual and the mean of the hourly activity intensities of all the individuals can be viewed in Figure \ref{raw}. It is evident that the activity of a single individual shows considerable variability where as the mean profile is fairly smooth indicating higher activity levels in the middle of the day. 

Using multivariate methods that treat observation at each time as a separate variable might pose statistical challenges to the analysis due to the temporal dependence present between them. Instead, viewing activity profile as a single entity will help capture this dependence and underlying activity changes such as acceleration, deceleration etc. Thus, we shall treat the activity data for each person as a function with domain being time of the day. It should be noted that it is not possible to observe the entire function over a continuum. The dataset contains PA intensity measurements $W_k, k=1,...,T$ at total $T$ time points for each individual. Smoothing methods \cite{ramsay2004functional} are used to recover the underlying function $W(\cdot)$ such that $W(t_k)=W_k,$ from the discrete observations. Functional Data Analysis (FDA) based techniques are increasingly used to analyze data sets with repeated measures including device based measures \cite{xu2019modeling,goldsmith2011penalized,sera2017using,leroux2019organizing}.


Despite it's advantages, accelerometers suffer from some limitations. The uniaxial accelerometers measure acceleration only along the vertical axis compared to the biaxial or the triaxial ones, which measure acceleration along multiple axes, thus giving more accurate measurements of activity \cite{corder2007accelerometers}.  Another major drawback of uniaxial accelerometers is that they cannot accurately measure nonambulatory activities such as cycling \cite{corder2007accelerometers, robertson2011utility}. Thus, the measured physical activity intensity is a proxy for the true PA intensity which is unmeasured. Considering this, the true pattern of PA intensity profile $X(t),t\in [0,1] $ is not directly observed and the activity intensity profile $W(t),$ is only a proxy for it. This can be modeled with a measurement error model. Thus, to obtain the estimate of the effect of the true activity intensity $X(\cdot)$ on BMI, we need a functional regression model that considers the problem of measurement error. Several existing  functional regression models \cite{yao2005functional,cardot2007smoothing,goldsmith2011penalized} allow for the possibility of measurement error at discrete realizations at which functions are observed  i.e., these methods assume $ W_k=X(t_k)+\widetilde{U}_k,$ where $W_k$ and $\widetilde{U}_k$ are observed data points and measurement error respectively. However,  $\widetilde{U}_1,...\widetilde{U}_k$ are assumed to be independent/uncorrelated. This assumption is fairly stringent particularly in the context of functional data. Method proposed in \cite{cai2015methods}, allows certain correlation structures for the error variables, however, their approach does not treat the error as a function. A functional linear model that allows a measurement error process, i.e. $W(t)=X(t)+\widetilde{U}(t)$, where $\widetilde{U}(\cdot)$ is the measurement error process is presented in \cite{chakraborty2017regression}. This, method requires for measurement error to be uncorrelated beyond a small interval of length $\delta,$ i.e $cov\{\widetilde{U}(t),\widetilde{U}(s)\}=0,|t-s|>\delta$ and it's performance is sensitive to this assumption \cite{jadhav2020functional}. Since this assumption may not hold for the  NHANES dataset we need another alternative. \cite{jadhav2020functional} developed a functional regression model allowing for a general measurement error process using replicates. Replicates are used to obtain estimate of the covariance of the error process $\widetilde{U}(\cdot).$ A concern with this approach is that the data over the 7-days needs to be divided into 2 sets, one set to obtain the daily average PA intensity profiles and the other set to estimate the error covariance function. This split will worsen the impact of the missing values on the daily average profiles and the covariance estimate of the error process. So, instead of splitting the data on activity intensity, we propose to  resolve the measurement error problem using the data on step counts which is recorded by the accelerometer in addition to the activity intensity.

Because step counts is also observed repeatedly over time, we treat it as a functional variable. Again, this data is averaged to obtain patterns of step count profiles over hourly intervals. Raw step count profile of randomly selected individuals and the mean step profile can be viewed in Figure \ref{raw}. This step-count functional variable is denoted as $Z(\cdot)$. We will use the step count as an instrument variable to account for measurement error. An instrument is generally required to satisfy the following assumptions (i) it must be correlated with true covariate $X(\cdot)$, (ii) it must be error free and independent of measurement error $\widetilde{U}(\cdot)$, and (iii) it must be a surrogate for the true activity intensity i.e., it must be independent of the response given $X(\cdot)$. 
The first assumption is reasonable. The third one is related to the assumptions in section 2.1 and will be discussed therein. 
The second one however, needs careful thought. The source of the measurement error in PA intensity arises from the lack of ability of uniaxial accelerometers to capture PA in non-vertical direction. However, walking is a relatively a simpler activity that we expect the accelerometers to capture fairly accurately. Moreover, the step counts are aggregated over 7 days. This aggregation helps deal with possible measurement errors in the step count data. In addition, different procedures are used to measure step counts and PA intensity. Thus, we assume that step counts do not contain any measurement error and they are independent from the measurement error in PA intensity. Step counts have been used as instruments for the related variable - energy expenditure in \cite{tekwe2019instrumental}. For additional details on the use of instruments in measurement error models  refer to \cite{carroll2006measurement}. \cite{tekwe2019instrumental} propose a functional regression where the relation between the functional instrument $Z(\cdot)$ and the unobserved variable of interest is assumed as $Z(t)=\delta X(t)+U(t),$ with $ U(\cdot)$ as the model error. Thus, the relation between function $X(\cdot)$ and $Z(\cdot)$ is constant over time. This, assumption may not hold for our purpose. Another concern is that functional methods addressing measurement error mainly focus on estimating the relation between the variables and do not provide a direct way to test whether they are related. We propose a novel method that overcomes the shortcomings of the available methods to produce a consistent estimate of the relationship between true PA intensity and BMI and a test that determines whether this relationship is significant.

\section{Methodology}
\label{s:model}

\subsection{Model}
We assume that the relation between $Y \in \mathbb{R}$ and the functional variable $X(\cdot)\in L^2[0,1] $ is the following functional linear model
\begin{align}\label{mm}
Y=\widetilde{\beta}_0+\int_0^1 \widetilde{\beta}(t)X(t)dt+e
\end{align}

The function $X(\cdot)$ is not directly observable and instead, variables $ W(\cdot), Z(\cdot)\in L^2[0,1] $ are observed such that the relation between the two functions follows the additive measurement error model 
\begin{align}\label{me}
W(t)=X(t)+\widetilde{U}(t),
\end{align}
where $\widetilde{U}(\cdot)$ is the measurement error process. The variable $W(\cdot) $ can be viewed as the ``contaminated" version of $X(\cdot).$ We assume that the data contains an instrument $ Z(\cdot) \in L^2[0,1]$ satisfying the following assumptions:\\
i)  $\{W(\cdot),Z(\cdot)\} $ is a surrogate for $X(\cdot)$, i.e. $E[Y \lvert \{X(\cdot),Z(\cdot),W(\cdot)\}]=E\{Y \lvert X(\cdot)\} ,$\\
ii) the relationship between $X(\cdot)$ and the instrument is the concurrent functional model 

\begin{align}\label{m1}
X(t)=\widetilde{\theta}(t)Z(t)+U(t)
\end{align} 
iii)  
the instrument $Z(\cdot) $ is independent of the errors $e,\widetilde{U}(\cdot),$ and $U(\cdot).$
Note that independence of $W(\cdot) $ and $X(\cdot) $ from model error $e$ implies assumption i).  In most cases independence is a reasonable assumption as long as variable and model choice is reasonable.

These conditions along with \eqref{mm}, \eqref{me}, and \eqref{m1} imply
\begin{align}\label{stp1}
E\{W(\cdot )\lvert Z(\cdot)\}=E\{X(\cdot )\lvert Z(\cdot) \},  
\end{align}
and 
\begin{align}\label{stp2}
E\{Y \lvert Z(\cdot )\}=\int_{0}^{1} \widetilde{\beta}_0+\widetilde{\beta}(t) E\{W(t) \lvert Z(t)\}dt=\widetilde{\beta}_0+\int_{0}^{1} \widetilde{\beta}(t) \widetilde{\theta}(t)Z(t)dt.
\end{align}



If the parameter function $\widetilde{\theta}(\cdot)$ is known, then the function $\widetilde{\beta}(\cdot)$ can be estimated using any one of the numerous techniques related to the functional linear model with scalar response such as \cite{muller2005generalized,reiss2017methods,kokoszka2017introduction}, where the response variable is $Y$ and the regressor variable is $V(t)=\widetilde{\theta}(t)\times Z(t)$. However, $\widetilde{\theta}(\cdot) $ is not known, so instead we use it's estimate in the following approximate model 
\begin{align} \label{apr}
E(Y|Z(\cdot)) \approx \widetilde{\beta}_0+\int_{0}^{1} \widetilde{\beta}(t) \widehat{V}(t)dt,
\end{align}
where $\widehat{V}(t)=\widehat{\widetilde{\theta}}(t)\times Z(t).$
\cite{carroll2006measurement} use a similar technique called regression calibration that involves replacing an unknown term in the model with its estimate to address measurement error for finite dimensional data.  We propose a two-step approach motivated by  \eqref{m1} and \eqref{apr} to obtain the estimate of the function $\widetilde{\beta}(\cdot)$. Before proceeding with the estimation, we briefly introduce the concept of basis expansion and accompanying notations that appear throughout this work.

\underline{Basis Expansion}: Let $f(\cdot)$ be a function in $L^2[0,1].$ It can be expanded as $f(t)=\sum_{i=1}^{\infty} f_i\phi_i(t)dt,$ where $\phi_i,\,i\geq 1$ denotes basis functions in $L^2[0,1]$ and coefficients $f_i=\int_{0}^{1} f(t)\phi_i(t)dt.$ We will denote the coefficients of a function $f(\cdot)$ as $f_i,\, i\geq 1$. A truncation strategy involves considering only a subset of all the components. These components are represented by a vector $ f_c=(f_1,...,f_p)',$ where the subscript in $f_c$ indicates that the vector contains coefficients from a basis expansion of a function $f(\cdot)$. 
Hence-onwards we will use $\phi_i(\cdot)$ and $\psi_i(\cdot)$ to denote different sets of basis functions in $L^2[0,1]$.

\subsection{Two-Step Estimation Algorithm }
Estimation of $\widetilde{\beta}(\cdot)$ involves two steps that result directly from \eqref{m1} and \eqref{apr}. In the first step we use \eqref{m1} to  obtain estimate of $\widetilde{\theta}(\cdot)$. In the second step, we use this estimate in \eqref{apr} to get the estimate of $\widetilde{\beta}(\cdot).$\\
\underline{Step 1}: This step involves estimating the function $\widetilde{\theta}(\cdot)$. There is significant literature on models where both the response and the regressor variables are functional \cite{chiou2004functional,maity2017nonparametric,kim2018additive,ramsay2004functional,benatia2017functional,ma2016dynamic}. These works use differing approaches, assumptions etc. For example, they use different penalization techniques, different functions spaces etc. The proposed approach uses the estimate of $\widetilde{\theta}(\cdot)$ to derive the estimate of $\widetilde{\beta}(\cdot)$. Hence, we need a unified framework of assumptions to study the asymptotic results for both of these models. Thus, we develop estimating procedure and asymptotics for the concurrent functional model. From the basis expansion,  $\widetilde{\theta}(t)=\sum_{j=1}^{\infty} \widetilde{\theta}_{j}\phi_j(t).$  From \eqref{stp1}, and  \eqref{m1} the relation between the instrument and the observed variable is $E\{W(t)|Z(t)\}=\sum_{j=1}^{\infty} \widetilde{\theta}_{j} \phi_j(t) Z(t)$. To address the infinitely many parameters, we use the following truncated model
\begin{align}\label{trun1}
E\{W(t)|Z(t)\} \approx \sum_{j=1}^{q} \widetilde{\theta}_{j} \phi_j(t) Z(t).
\end{align}
The truncation is not restrictive as it is assumed that $q\to \infty$. Thus, the usual multivariate methods cannot be used as the number of parameters diverges. Let $ \widetilde{\theta}_c=(\widetilde{\theta}_1,...,\widetilde{\theta}_q)^{'} $ and $  \phi(t)=\{\phi_1(t),...,\phi_q(t)\}^{'}.$ Given identical and independent observations on $(Y_i,X_i,Z_i),\,i=1,...n$ of $Y, W(\cdot)$ and $Z(\cdot)$, the estimate of the parameter $ \widetilde{\theta}_c $ is

$$\underset{\theta \in R^q}{argmin} \sum_{i=1}^{n}\int \left[W_i(t)- 
Z_i(t)\theta^{'}\phi(t) \right]^2dt    $$

Alternatively, this solution can be characterized as the solution of estimating equation 

\begin{align}\label{SM1_1}
S(\theta):=\sum_{i=1}^{n} \int W_i(t)Z_i(t)\phi(t)dt-\int Z_i^2(t)\phi(t)\phi^{'}(t)\theta dt=0.
\end{align}
Thus, $\widehat{\widetilde{\theta}}_c=\left(\sum_{i=1}^{n}\int Z_i^2(t)\phi(t)\phi^{'}(t)dt \right)^{-1}\sum_{i=1}^{n} \int W_i(t)Z_i(t)\phi(t)dt, $ which is used to obtain $\widehat{\widetilde{\theta}}(t)=\phi(t)^{'}\widehat{\widetilde{\theta}}_c .$	Recall that $V(t)=\widetilde{\theta}(t)\times Z(t)$ and  $\widehat{V}(t)=\widehat{\widetilde{\theta}}(t)\times Z(t).$ With this estimate of $\widehat{V}(\cdot)$ proceed to the second step. 

\underline{Step 2}: This step results from $E\{Y \lvert Z(t)\}=\widetilde{\beta}_0+\int_{0}^{1} \widetilde{\beta}(t)V(t)dt,$ which was obtained in \eqref{apr}. Using basis expansions $V(t)= \sum_{j=1}^{\infty} V_j\psi_j(t)$ and $\widetilde{\beta}(t)=\sum_{j=1}^{\infty} \widetilde{\beta}_j\psi_j(t)$ we obtain $E\{Y \lvert Z(\cdot)\}=\widetilde{\beta}_0+\sum_{j=1}^{\infty} \widetilde{\beta}_jV_j.$ Truncation of this model leads to $E\{Y \lvert Z(\cdot)\}=\widetilde{\beta}_0+\sum_{j=1}^{p} \widetilde{\beta}_jV_j.$ Again, this truncation is not restrictive as we allow $p \to \infty.$  The estimate of $\widetilde{\beta}_0$ and $\widetilde{\beta}_c=(\widetilde{\beta}_1,...\widetilde{\beta}_p)^{'}$ can be easily obtained if the function $V$ and hence, it's coefficients $V_c=(V_1,...V_p)$ are known. Though $V(\cdot)$ is unknown, it's estimate $\widehat{V}(\cdot) $ and it's coefficient vector $\widehat{V}_c$ are available. For expediency, denote $\widetilde{\beta}_c=(\widetilde{\beta}_0,\widetilde{\beta}_1,...\widetilde{\beta}_p)^{'}$ and $V_c=(V_0,V_1,...V_p)^{'}, V_0=1.$ Given identical and independent observations on $Y_i$ of $Y$, the estimate of $\widetilde{\beta}_c$  is the minimizer of the following:

\begin{align}
\underset{\beta \in R^{p+1}}{argmin} \sum_{i=1}^{n}\left(Y_i-\beta^{'}\widehat{V}_{ci}\right)^2
\end{align}
The corresponding estimating equation is
\begin{align}\label{score}
\widehat{U}(\beta)= \sum_{i=1}^{n} (Y_i-\beta^{'}\widehat{V}_{ci})\widehat{V}_{ci}^{'}=0
\end{align}

The solution of this is $\widehat{\widetilde{\beta}}_c=(\sum_{i=1}^{n}\widehat{V}_{ci}\widehat{V}_{ci}^{'})^{-1}(\sum_{i=1}^{n}Y_i\widehat{V}_{ci}) $ and $\widehat{\widetilde{\beta}}(t)=\psi(t)^{'}\widehat{\widetilde{\beta}}_c .$

\begin{remark}
	A more general model than the concurrent one  to capture the relation between  $X(\cdot)$ and $Z(\cdot)$ is $ X(t)=\int_s \alpha(s,t)Z(s)ds+U(t).$ Truncated basis expansion yields  $\alpha(s,t)=\sum_{k=1}^{K}$ $\sum_{m=1}^{M} \alpha_{km}\phi_k(t)\psi_m(s).$	Then, the approximate model for function $X(\cdot)$ is $X(t)= \sum_{k=1}^{K}$ $\sum_{m=1}^{M} \alpha_{km}\int_{s}\phi_k(t)\psi_m(s) Z(s)ds+U(t).$ Let $g_i(t)=\{\int  Z_i(s)\phi_1(t)\psi_1(s)ds,...,\int_s Z_i(s)\phi_K(t)\psi_M(s)ds\}$ and let $\alpha_c=(\alpha_{11},...,\alpha_{KM}).$ 
	The parameters $\alpha_{c}$ is estimated by minimizing $$\underset{\alpha \in R^{KM}}{argmin} \sum_{i=1}^{n}\int_t \left[W_i(t)- \alpha{'} g_i(t)\right]^2dt.$$ Thus, there will be $KM$ estimating equations instead of $q$. The main difference between this model and the concurrent one is the number of parameters. The estimating procedure and even the asymptotics do not change fundamentally.  Hence, for convenience, we use the concurrent model though our methodology is  valid for the more general one.  
\end{remark}

\begin{remark}
	We may choose different basis functions for $\widetilde{\beta}$ and $V.$ We use the same basis function for convenience. For applications, we recommend using Bsplines due to their flexibility in modeling different types of curves or use basis obtained from the covariance function of $V(\cdot)$ to obtain a relatively frugal representation in the basis expansion. We have used the latter option in the subsequent analyses .
\end{remark}

\subsection{Hypothesis Test}

Once the relationship between $Y$ and $ X(\cdot)$ i.e. $\widehat{\widetilde{\beta}}_c$ is determined, we proceed to investigate it's significance via the null hypothesis $H^*_0:\widetilde{\beta}(\cdot)=0$ vs $ H^*_1:  \widetilde{\beta}(\cdot) \neq 0.$ This is equivalent to testing $H_0: \widetilde{\beta}_c=(0,...0)$ vs. the alternative that $H_0$ is not true. Let $\widehat{\Gamma}=\left\{\dfrac{\sum_{i=1}^{n}\widehat{V}_{ci}\widehat{V}^{'}_{ci}}{\widehat{var}(Y)n}\right\} $. In order to test this hypothesis we propose the test statistic 
\begin{align}\label{testst}
\widehat{T}=\dfrac{n\widehat{\widetilde{\beta}}'\widehat{\Gamma}\widehat{\widetilde{\beta}}-(p+1)}{\sqrt{2(p+1)}}.
\end{align}

Similar form of test statistic can be found in \cite{muller2005generalized} and \cite{jadhav2017dependent}, though these works do not consider measurement error. We will show in the next section that asymptotically and under $H_0$, $\widehat{T}$ has a standard normal distribution and thus p-value for the test can be easily determined. This also leads to the following confidence intervals  \cite{muller2005generalized}. Consider $\widetilde{\beta}(t) = \sum_{j=1}^{p} \widetilde{\beta}_j \psi_j(t), $ where the basis functions are orthonormal. Let $c(\alpha) = \left(p+1+\sqrt{2(p+1)}\phi(1-\alpha)\right)/n, $ where, $\phi(1-\alpha)$ is $100(1-\alpha)$th percentile of the standard normal distribution. Let $(e_i,\lambda_i), i=1,...,p+1 $ be the eigenvectors and eigenvalues and of $ \widehat{\Gamma}$ with $e_k = (e_{k1},...,e_{kp+1})'. $ Denote, $\omega_k(t) = \sum_{l=1}^{p+1} \psi_l(t)e_{kl}. $ Then, the asymptotic, approximate $(1-\alpha) $ confidence band is given as

\begin{align}\label{ci}
\widehat{\widetilde{\beta}} \mp \sqrt{c(\alpha)\sum_{j=1}^{p}\dfrac{\omega_j(t)^2}{\lambda_k} }.
\end{align}
\section{Asymptotic Properties}

In this section, we list results that establish the weak consistency of the suggested estimates, the asymptotic distribution of the test statistic in \eqref{testst}, and the necessary assumptions. The norm of the function $ f(\cdot)\in L^2[0,1] $ is  $\|f\|^2_L=\int_0^1 f^2(t)dt,$ while the Frobenius norm of a vector or a matrix is denoted as $\| \cdot \|. $ The minimum and maximum eigenvalue of a matrix $A$ are expressed as $\lambda_{min}(A) $ and $\lambda_{max}(A) $ respectively. All integrals are taken over the interval $[0,1].$

\begin{assumption}\label{A0}
	The number of parameters diverge i.e.
	$p,q\to \infty $ as $n \to \infty.$ 
\end{assumption}

\begin{assumption}\label{A1}
	The error variables $e_i,U_i(\cdot)$ and $ \widetilde{U}(\cdot)$ are centered and the following moments are bounded
	$$ \text{max}\{ \|\widetilde{\theta}(\cdot)\|_L, \|\widetilde{\beta}(\cdot)\|_L,E\|Z(\cdot)\|_L^4 ,\, E\|U(\cdot)\|_L^4 ,\,E\|\widetilde{U}(\cdot)\|_L^4, E\|X(\cdot)\|_L^4, E(e^2)\}< \infty.  $$
\end{assumption}

\begin{assumption}\label{A2}
	The rate of divergence of $q$ is restrained, i.e. $ \sqrt{q/n} \to 0 $ as $n \to \infty$.
\end{assumption}

\begin{assumption}\label{A3}
	There exist two positive constants $b_1$ and $b_2$ such that $$b_1< \lambda_{min} \left[n^{-1}\sum_{i=1}^{n}\int Z_i^2(t)\phi(t)\phi^{'}(t)dt\right] \leq \lambda_{max} \left[n^{-1}\sum_{i=1}^{n}\int Z_i^2(t)\phi(t)\phi^{'}(t)dt\right] < b_2 .$$
\end{assumption}

Assumptions in \ref{A1} and \ref{A2} are required to obtain consistent estimates of various fundamental parameters involved in the model such as the covariance function, its eigenvalues and eigenvectors. Functional Data Analysis (FDA) literature contains several works \cite{muller2005generalized} with assumptions similar to \ref{A2} that limit the growth of the parameter dimension. Assumption \ref{A3} requires sample covariance matrix to have bounded, positive eigenvalues. All these assumptions are routine and are typically satisfied by the data. 

\begin{theorem}\label{thm1}
	Under Assumptions \ref{A1} - \ref{A3} solution $\widehat{\widetilde{\theta}}_c$ of \eqref{SM1_1} is weakly consistent i.e.
	$$ \| \widehat{\widetilde{\theta}}_c - \widetilde{\theta}_c\|=O_p(\sqrt{q/n}) =o_p(1),  $$	
	
\end{theorem}	
This theorem implies that the function estimate $\widehat{\widetilde{\theta}}(\cdot)  $ is consistent.

We next state assumptions needed to establish the consistency of the estimate of the function $\widetilde{\beta}(\cdot).$

\begin{assumption}\label{A4}
	The instruments	$Z_i,\, i=1,..n$ are uniformly bounded.
\end{assumption}

\begin{assumption}\label{A5}
	Growth of the number of parameters $p$ is restricted as
	$$\sqrt{p/n} \to 0.  $$
\end{assumption}

\begin{assumption}\label{A7}	
	
	There exists a positive constant $b_3$ and $b_4
	$ such that 
	$$b_3<\lambda_{min}\left(n^{-1}\sum_{i=1}^{n}\widehat{V}_{ci}\widehat{V}_{ci}^{'}\right)\leq \lambda_{max}\left(n^{-1}\sum_{i=1}^{n}\widehat{V}_{ci}\widehat{V}_{ci}^{'}\right)<b_4$$
	
\end{assumption}

\begin{theorem}\label{thm2}
	Under the Assumption  \ref{A1} and Assumptions \ref{A4} - \ref{A7}, the estimate $\widehat{\widetilde{\beta}}_c$ from \eqref{score} is weakly consistent i.e.
	$$ \| \widehat{\widetilde{\beta}}_c - \widetilde{\beta}_c\| = O_p(\sqrt{p/n}) . $$	
\end{theorem}

This also implies the consistency $\widehat{\widetilde{\beta}}(\cdot)$. 
The following additional are assumptions similar to those in \cite{jadhav2017dependent}, are needed to prove that the asymptotic distribution of the statistic $\widehat{T}$ in \eqref{testst} is standard normal.

\begin{assumption}\label{A8}
	
	Assume that $pn^{-1/6}\to 0  $
\end{assumption}

\begin{assumption}\label{A9}
	Then, eigenvalues of $ \Gamma = E(\widehat{\Gamma})$ are bounded and $ \|\widehat{\Gamma}^{-1}\|<O_p(p^{1/2}) $
\end{assumption}

\begin{theorem}\label{thm3}
	Under the Assumptions \ref{A0} - \ref{A9},
	$$T=\dfrac{n(\widehat{\widetilde{\beta}}-\widetilde{\beta})'\Gamma(\widehat{\widetilde{\beta}}-\widetilde{\beta})-(p_n+1)}{\sqrt{2(p_n+1)}}\to N(0,1)$$	
	
	$$\widehat{T}=\dfrac{n(\widehat{\widetilde{\beta}}-\widetilde{\beta})'\widehat{\Gamma}(\widehat{\widetilde{\beta}}-\widetilde{\beta})-(p_n+1)}{\sqrt{2(p_n+1)}}\to N(0,1)$$
\end{theorem}	

The proofs of Theorem \ref{thm1} and Theorem \ref{thm2} are given in the Web Appendix. Since, the proof \ref{thm3} is along the same lines as Theorem 4.1 from \cite{muller2005generalized}, and Theorem 2 from \cite{jadhav2017dependent}, we do not repeat it. 

The proofs of Theorem \ref{thm1} and Theorem \ref{thm2} are given in the Web Appendix. Since, the proof \ref{thm3} is along the same lines as Theorem 4.1 from \cite{muller2005generalized}, and Theorem 2 from \cite{jadhav2017dependent}, we do not repeat it. 

\section{Simulation Study}

\subsection{Estimation}

We study the properties of the proposed estimation method and compare it's performance with alternative methods.  In the following, $N(a,b)$ denotes a normal distribution with mean $a$ and variance $b$.

Let $\phi_z,\,\phi_{\theta},\,\phi_{\beta}$ denote basis functions in $L^2[0,1].$ An identically distributed and independent sample of instruments $Z_i(\cdot)$ are generated as $Z_i(t)=\sum_{j=1}^{k_0} z_{ij}\phi_{zj}(t),\,i=1,...n,\, z_{ij}\overset{i.i.d}{\sim} N(0,1)  $. Note that Assumption \ref{A4} assumes that the instruments are bounded. To show that the  proposed method works with a moderate violation of this assumption  we do not generate bounded instruments. Parameter functions $\widetilde{\theta}(\cdot) $ and  $\widetilde{\beta}(\cdot)$ are generated as $\widetilde{\theta}(t)=\sum_{j=1}^{q_0} a_j\phi_{\theta j}(t),\, a_j=j/q_0$ and $ \widetilde{\beta}(t)=\sum_{j=1}^{p_0} b_j\phi_{\beta j}(t),\, b_j=1/j.  $ The instruments are used to obtain the true functions $X_i(\cdot)$ as $X_i(t)=\widetilde{\theta}(t)Z_i(t)+U_i(t),$ where the error process $U_i(\cdot),$ is a Brownian Motion. Note that functions $X_i(\cdot)$ are treated as unobserved. Let  $e_i\overset{i.i.d}{\sim} N(0,0.1)$, then the response is $Y_i=\int \widetilde{\beta}(t) X_i(t)dt  $ $+ e_i$. 
There are two main characteristics of the measurement error process that can impact the accuracy of the estimates. We consider them in the following  two scenarios. \\
\underline{Scenario 1}: In this scenario we study the impact of the range of measurement error. Specifically, we generate a Gaussian Process $\widetilde{U}_i(\cdot) $ with the covariance function $(0.1)\text{exp}(-(s-t)^2/(2l^2)).$ This is a squared exponential function where the covariance depends on the distance between the points. Small values of $l$ reduce the covariance between different points and especially those that are further away. Thus, smaller values of $l$ lower the range of dependence and larger values increase it.\\
\underline{Scenario 2}: Here, we keep the range of dependence constant and vary the amount of measurement error i.e. we generate a Gaussian Process $\widetilde{U}_i(\cdot) $ with the covariance function $\sigma\text{exp}(-(s-t)^2/(2(0.05)^2))$. The parameter $\sigma$ controls the amount of measurement error with larger values leading to larger error.\\ 
The proxy is created as $W_i(t)=X_i(t)+\widetilde{U}_i(t).$  For comparison, we use alternative methods for estimation. To implement the proposed method, in \eqref{SM1_1} of Step 1, we use the B-spline basis for $\widetilde{\theta}(
\cdot)$. The number of bases or alternatively the number of parameters $q$ are selected using 5 fold cross-validation from the values $4, 6, 8, 10.$ Note that these values are not too large or too small ensuring that we adequately capture the functions with out over-parameterizing. Functional methods are not very sensitive to the number of basis functions as long we choose enough bases to capture the data characteristics.  For Step 2 in \eqref{score}, we use basis obtained from the covariance function of $\widehat{V}_i(\cdot)$ i.e. we employ functional principal component analysis which ensures that the number of parameters $p$ is adequately small. Again 5-fold cross-validation is used used to choose an appropriate value for $p$. 

The naive estimate which ignores the measurement error is implemented using method from \cite{muller2005generalized}. Let $W= (W_{1c}^{'},...,W_{nc}^{'})^{'}$ be the Fourier coefficients obtained from $ W_i(\cdot), i=1,...,n$ and $Y=(Y_1,...Y_n)^{'}$. Then, the naive estimate denoted as $ \widehat{\beta}_{\text{naive}}=(W^{'}W)^{-1}W^{'}Y.$  Alt.1 is the method suggested in Tekwe et al. (2019) by assuming that $Z(t)=\delta X(t)+U(t)$. The method suggested by Chakraborty and Panaretos (2017) which assumes that the range of dependence for the measurement error process is small is referred to as Alt.2. Tables \ref{one} and \ref{two} report the average estimation error of the various estimates of  $\widetilde{\beta}(t) $ based on 500 replicates for all the methods along with the average estimation error for $\widehat{\widetilde{\theta}}(t) $. The estimation error for an estimate $ a(\cdot) $ is calculated as $\int \{a(t)-\widetilde{\beta}(t)\}^2dt.$ The  average value of the $q$ and $p$ obtained from the cross-validation is also reported.The proposed method is referred to as Prop. in the Tables. For the basis functions we choose $\phi_z$ as B-spline, $\phi_{\beta}$ as Fourier basis and $\phi_{\theta}$ as monomial basis with $k_0=5,\, q_0=3 $ and $p_0=3.$ To investigate the properties of the proposed method we vary the values of $n, \sigma$ and $l$. Tables \ref{one_var} and \ref{two_var} report the variance estimation errors of the various estimates.

\begin{center}
	\begin{table}[t]%
		\centering
		\caption{Scenario 1 - n is the sample size, $l$ controls the range of dependence of measurement error , $\theta_{er}$ is the average estimation error of $\widehat{\widetilde{\theta}}. $ Prop, naive, Alt.1 and Alt.2 denote the average estimation error for proposed, naive, Alt.1 and Alt.2 methods respectively. $\bar{q} $ and $\bar{p} $ is the average values of $p$ and $q$ from the cross-validation. \label{one}}%
		\begin{tabular*}{500pt}{@{\extracolsep\fill}llccccccc@{\extracolsep\fill}}
			\toprule
			\textbf{$n$} & \textbf{$l$}  & \textbf{$\theta_{er}$}  & \textbf{Prop}  & \textbf{Naive} & \textbf{Alt.1}& \textbf{Alt.2}& $\bar{q} $& $\bar{p} $ \\
			\midrule
500	&	0.05	& 1.85E-04 &	0.096	&	0.382	&	0.212	&	0.243	&	4.804	&	4.996	\\
1000	&	0.05	&	9.14E-05	&	0.042	&	0.258	&	0.160	&	0.185	&	4.756	&	5	\\
3000	&	0.05	&	3.2E-05	&	0.015	&	0.163	&	0.126	&	0.130	&	4.776	&	5	\\
500	&	0.1	&	2.53E-04	&	0.083	&	0.770	&	0.197	&	0.236	&	4.600	&	5	\\
1000	&	0.1	&	1.29E-04	&	0.042	&	0.702	&	0.146	&	0.177	&	4.512	&	5	\\
3000	&	0.1	&	4.37E-05	&	0.015	&	0.958	&	0.117	&	0.136	&	4.540	&	5	\\
500	&	0.5	&	3.57E-05	&	0.089	&	0.749	&	0.256	&	0.245	&	4.620	&	5	\\
1000	&	0.5	&	1.65E-04	&	0.045	&	0.718	&	0.184	&	0.174	&	4.536	&	5	\\
3000	&	0.5	&	5.87E-05	&	0.015	&	0.691	&	0.132	&	0.130	&	4.708	&	5	\\
			\bottomrule
		\end{tabular*}
	\end{table}
\end{center}

\begin{center}
	\begin{table}[t]%
		\centering
		\caption{Scenario 1 - n is the sample size, $l$ controls the range of dependence of measurement error , $\theta_{er}$ is the variance of the estimation error of $\widehat{\widetilde{\theta}}. $ Prop, naive, Alt.1 and Alt.2 denote the variance of the estimation error for proposed, naive, Alt.1 and Alt.2 methods respectively.\label{one_var}}%
		\begin{tabular*}{500pt}{@{\extracolsep\fill}llccccc@{\extracolsep\fill}}
			\toprule
			\textbf{$n$} & \textbf{$l$}  & \textbf{$\theta_{er}$}  & \textbf{Prop}  & \textbf{Naive} & \textbf{Alt.1}& \textbf{Alt.2} \\
			\midrule
500	&	0.05	&	2.20E-08	&	0.0093	&	0.4642	&	0.0346	&	0.0274	\\
1000	&	0.05	&	5.51E-09	&	0.0014	&	0.1759	&	0.0116	&	0.0121	\\
3000	&	0.05	&	6.28E-10	&	0.0002	&	0.0274	&	0.0030	&	0.0032	\\
500	&	0.1	&	4.05E-08	&	0.0069	&	0.2349	&	0.0278	&	0.0307	\\
1000	&	0.1	&	1.02E-08	&	0.0016	&	0.4780	&	0.0106	&	0.0110	\\
3000	&	0.1	&	1.16E-09	&	0.0002	&	0.5683	&	0.0032	&	0.0034	\\
500	&	0.5	&	8.68E-08	&	0.0083	&	0.0677	&	0.0380	&	0.0331	\\
1000	&	0.5	&	2.05E-08	&	0.0018	&	0.0306	&	0.0120	&	0.0106	\\
3000	&	0.5	&	2.26E-09	&	0.0002	&	0.0094	&	0.0027	&	0.0027	\\
			\bottomrule
		\end{tabular*}
	\end{table}
\end{center}

From the results in Table \ref{one}, we can see that the estimation error of $\widehat{\widetilde{\theta}}(\cdot)$ is quite low for all settings. The proposed estimator performs significantly better than all the alternative methods even when the range of dependence is increased (larger $l$). From Table \ref{one_var} the error variance of the proposed method is lower than all than alternatives. The error seems to decreases with the sample size for all the methods as expected. The Naive methods show slight increase in the error as we increase $l$ indicating that the amount of dependence in the error may be an important factor.

\begin{center}
	\begin{table}[t]%
		\centering
		\caption{Scenario 2 - n is the sample size, $\sigma$ denotes measurement error variance, $\widehat{\widetilde{\theta}}_{er}$ is the average estimation error of $\widehat{\widetilde{\theta}}. $ Prop, Naive, Alt.1 and Alt.2 denote the average estimation error for proposed, naive, Alt.1 and Alt.2 methods respectively. $\bar{q} $ and $\bar{p} $ is the average values of $p$ and $q$ from the cross-validation. \label{two}}%
		\begin{tabular*}{500pt}{@{\extracolsep\fill}llccccccc@{\extracolsep\fill}}
			\toprule
			\textbf{$n$} & \textbf{$\sigma$}  & \textbf{$\theta_{er}$}  & \textbf{Prop}  & \textbf{Naive} & \textbf{Alt.1}& \textbf{Alt.2}& $\bar{q}_{est} $& $\bar{p}_{est} $  \\
			\midrule
500	&	0.1	&	1.75E-04	&	0.085	&	0.409	&	0.218	&	0.258	&	4.616	&	5	\\
1000	&	0.1	&	9.21E-05	&	0.044	&	0.246	&	0.167	&	0.190	&	4.668	&	5	\\
3000	&	0.1	&	2.93E-05	&	0.016	&	0.174	&	0.125	&	0.130	&	4.676	&	5	\\
500	&	0.5	&	8.85E-04&	0.093	&	0.606	&	0.217	&	0.300	&	4.752	&	5	\\
1000	&	0.5	&	4.60E-04	&	0.051	&	0.541	&	0.156	&	0.248	&	4.832	&	5	\\
3000	&	0.5	&	1.47E-04	&	0.016	&	0.534	&	0.116	&	0.183	&	4.708	&	5	\\
500	&	1	&	1.79E-03	&	0.107	&	0.762	&	0.279	&	0.381	&	4.728	&	5	\\
1000	&	1	&	9.14E-04	&	0.052	&	0.732	&	0.175	&	0.287	&	4.800	&	5	\\
3000	&	1	&	2.97E-04	&	0.018	&	0.716	&	0.117	&	0.201	&	4.748	&	5	\\	
		\bottomrule
		\end{tabular*}
	\end{table}
\end{center}

\begin{center}
	\begin{table}[t]%
		\centering
		\caption{Scenario 2 - n is the sample size, $\sigma$ denotes measurement error variance, $\widehat{\widetilde{\theta}}_{er}$ is the variance of estimation error of $\widehat{\widetilde{\theta}}. $ Prop, Naive, Alt.1 and Alt.2 denote the variance of the estimation error for proposed, naive, Alt.1 and Alt.2 methods respectively.\label{two_var}}%
		\begin{tabular*}{500pt}{@{\extracolsep\fill}llccccc@{\extracolsep\fill}}
			\toprule
			\textbf{$n$} & \textbf{$\sigma$}  & \textbf{$\theta_{er}$}  & \textbf{Prop}  & \textbf{Naive} & \textbf{Alt.1}& \textbf{Alt.2}\\
			\midrule
500	&	0.1	&	2.15E-08	&	0.0076	&	0.4970	&	0.0343	&	0.0367	\\
1000	&	0.1	&	5.64E-09	&	0.0018	&	0.1335	&	0.0121	&	0.0114	\\
3000	&	0.1	&	5.96E-10	&	0.0002	&	0.0276	&	0.0028	&	0.0027	\\
500	&	0.5	&	5.29E-07	&	0.0082	&	0.0715	&	0.0579	&	0.0550	\\
1000	&	0.5	&	1.47E-07	&	0.0020	&	0.0262	&	0.0194	&	0.0247	\\
3000	&	0.5	&	1.59E-08	&	0.0002	&	0.0092	&	0.0047	&	0.0071	\\
500	&	1	&	2.14E-06	&	0.0115	&	0.0302	&	0.0992	&	0.1084	\\
1000	&	1	&	6.03E-07	&	0.0038	&	0.0115	&	0.0273	&	0.0376	\\
3000	&	1	&	6.17E-08	&	0.0003	&	0.0044	&	0.0068	&	0.0093	\\	
	\bottomrule
		\end{tabular*}
	\end{table}
\end{center}

From Table \ref{two} we observe the effects of increasing the amount of measurement error $(\sigma)$. The proposed method performs significantly better than the alternatives. Table \ref{two_var} shows that the error variance of the proposed estimator is quite low and lower than the alternatives. Again, the the error decreases as sample size increases for all the methods.  The error of the Naive estimate shows a clear increase with the increase in measurement error, as expected. Alt.1 and Alt.2 are not affected by this increase. Overall, in both scenarios our method is able to account for measurement error quite well.

\begin{figure}[t]
	\centerline{\includegraphics[width=3.5in,height=4in]{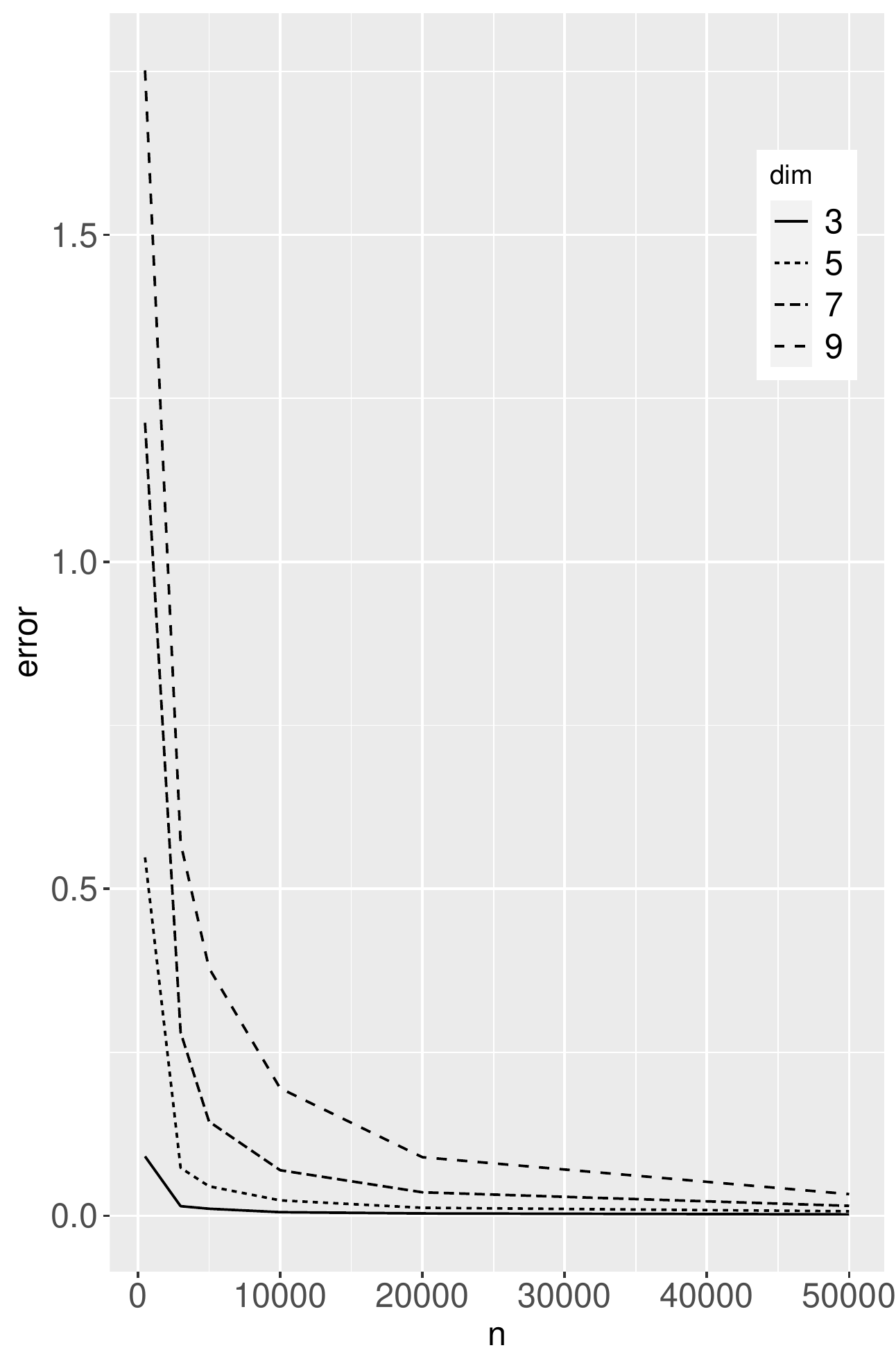}}
	\caption{This figure shows the estimation error as a function of the sample size $n$. dim indicates the dimension of the parameters in the study. dim of 3  denotes $p_0=3$ and $q_0=3.$ \label{con}}
\end{figure}

To demonstrate the asymptotic properties of the proposed method, we examine the estimation error while increasing the dimensions of the parameter functions along with the sample size. Data was generated in the same manner as above with $\sigma=0.1, l=0.05$. The number of basis functions  i.e. $p_0$ and $q_0$ are varied along with the sample size $n$. Figure \ref{con} shows that even if we increase the parameter function dimensions (denoted as dim), the estimation error will lower as the  sample size increases  in accordance with the asymptotic properties,
 
\subsection{Inference}
We will now study the proposed test procedure based on Theorem \ref{thm3} to determine whether $\widetilde{\beta}(t)=0$ or not. Functional data is generated the same way as in the Estimation section. We generate the measurement error process from the covariance function $0.1\text{exp}(-(s-t)^2/(2(0.05)^2)).$ For the response, we use $ Y_i=\delta \int \widetilde{\beta}(t)X_i(t)+e_i,$ where all the components are the same as before. We use $\delta$ to control the magnitude of the effect of the function $ \widetilde{\beta}(\cdot)$ on $Y$. Larger value of $\delta$ leads to stronger association and $\delta=0$ leads to no association between $Y$ and $X(\cdot).$ For comparison we also conduct the test using the naive estimate $\widehat{\beta}_{naive}.$ For this , we use the test statistic 
\begin{align}\label{naivetest}
T_w=\dfrac{n\widehat{\beta}_{naive}^{'}\Gamma_w\widehat{\beta}_{naive}-p}{\sqrt{2p}},
\end{align}
where $\Gamma_w=W^{'}W/n$. Since the statistic $T_w$ will have an asymptotic standard normal distribution in the absence of measurement error, we compute the p-value using the standard normal distribution. We refer to this test as the naive test. Table \ref{three} reports the empirical power and Type one error for the proposed test as well as the naive test based on 1000 replicates and level $\alpha=0.05$. 

\begin{center}
	\begin{table}[t]%
		\centering
		\caption{n is the sample size, $\delta$ indicates the strength of the relationship between $Y$ and $X(\cdot)$, Prop, Naive denote the empirical power of the proposed test and naive test respectively.\label{three}}%
		\begin{tabular*}{500pt}{@{\extracolsep\fill}llcc@{\extracolsep\fill}}
			\toprule
			\textbf{$n$} & \textbf{$\delta$}  & \textbf{Prop}  & \textbf{Naive}  \\
			\midrule
			200	&	0	&	0.070	&	0.106	\\
			500	&	0	&	0.066	&	0.12	\\
			1000	&	0	&	0.058	&	0.104	\\
			3000	&	0	&	0.049	&	0.101	\\
			200	&	0.1	&	0.106	&	0.118	\\
			500	&	0.1	&	0.176	&	0.23	\\
			1000	&	0.1	&	0.311	&	0.352	\\
			3000	&	0.1	&	0.741	&	0.748	\\
			200	&	0.3	&	0.326	&	0.358	\\
			500	&	0.3	&	0.905	&	0.891	\\
			1000	&	0.3	&	0.994	&	0.993	\\
			3000	&	0.3	&	1	&	1	\\ 
			\bottomrule
		\end{tabular*}
	\end{table}
\end{center}

From Table \ref{three}  (values corresponding to $\delta=0$) we can see that Type-I error is controlled for the proposed test but not the naive one. This is expected as the asymptotic distribution of $T_w$ is not  standard normal due to measurement error. For the proposed test, power increases with sample size and the effect size $\delta$.

Thus, from the simulation results we observe that the proposed method provides better estimates than the naive method and other alternative methods.
\section{Data Application}

\textbf{NHANES: Data problem background:}

\begin{figure}[t]
	\centerline{\includegraphics[width=4in,height=3.5in]{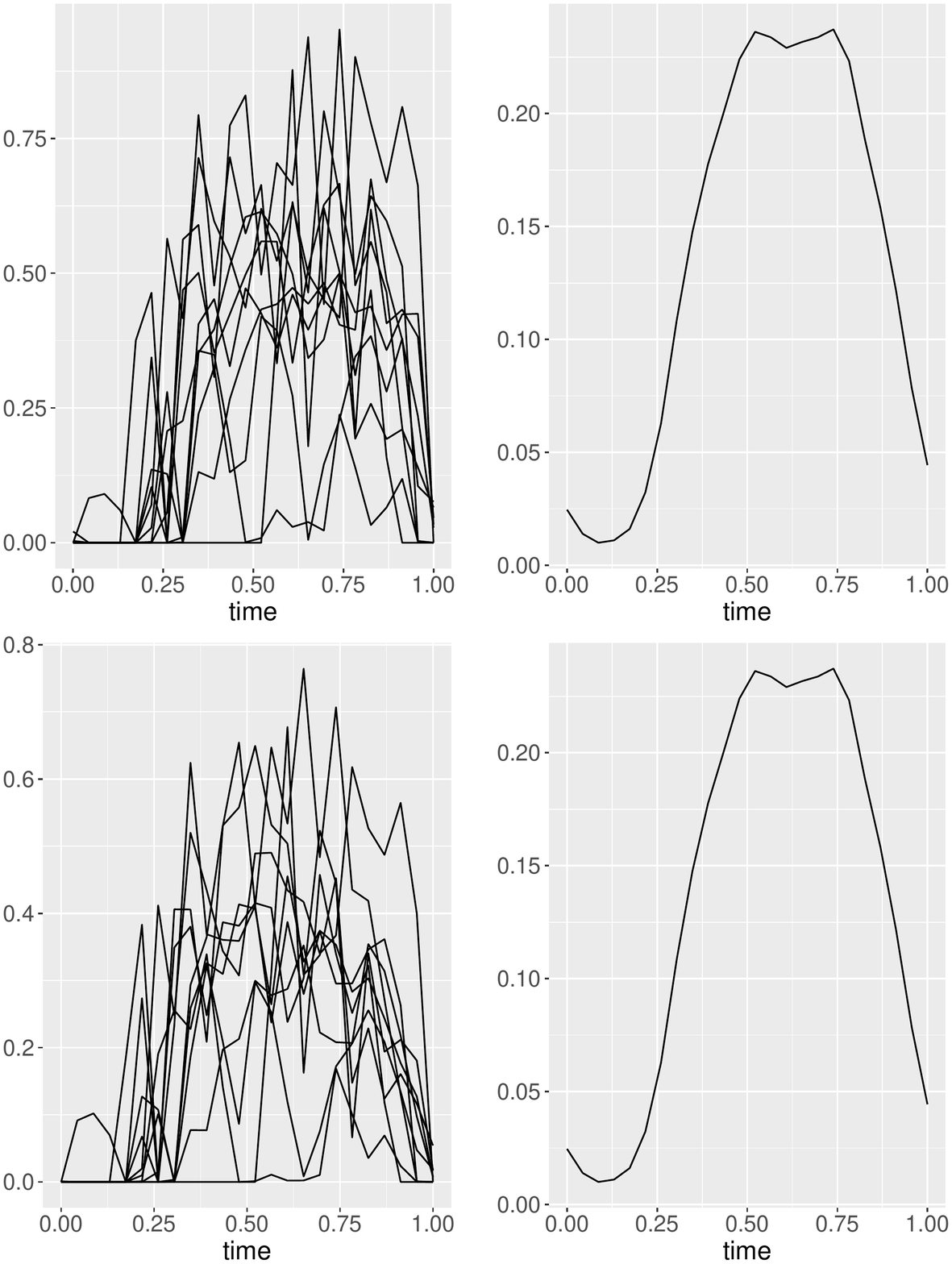}}
	\caption{Top row left panel contains the raw physical activity intensity and the right panel depicts it's mean. Bottom left panel has the raw step count data and the right one contains the mean of this data.\label{raw}}
\end{figure}

In this section, we apply our method to the motivating example of the NHANES data set. In this application, 1900 adults aged 20 and above who were interviewed in the 2005-2006 cycle are included from the NHANES data base. Prior to analyzing the data, we applied sample weights to account for the oversampling of racial groups using analytic guidelines provided for the NHANES \cite{johnson2013national} data. Since the magnitude of the observations is very large, we standardize the data by dividing with a constant so that all observations are less than 1. This transformation preserves the underlying patterns and ensures that the data and associated moments are bounded (see Assumptions \ref{A1}, \ref{A3}, \ref{A4}).  Additionally, time is scaled such that $t=0$ indicates start of the day and $t=1$ indicates it's end. 
Figure \ref{raw} displays standardized data of 10 randomly picked individuals along with the overall mean.

Though individual curves are erratic, the mean profiles reveal an underlying pattern where an average person is not very active at the start of the day that includes sleeping hours, followed by increasing activity levels which drop off at the end of the day. To capture the pattern over the noise, we apply smoothing techniques \cite{ramsay2004functional} to the data set. We use the B-spline basis and penalize the second derivatives to ensure smooth functions. This is executed using the Data2fd function from R package fda which converts the observed discrete data to a functional one with the specified degree of smoothness. This, smoothened data can be visualized in Figure~\ref{smooth} below.

\begin{figure}[t]
	\centerline{\includegraphics[width=4in,height=3.5in]{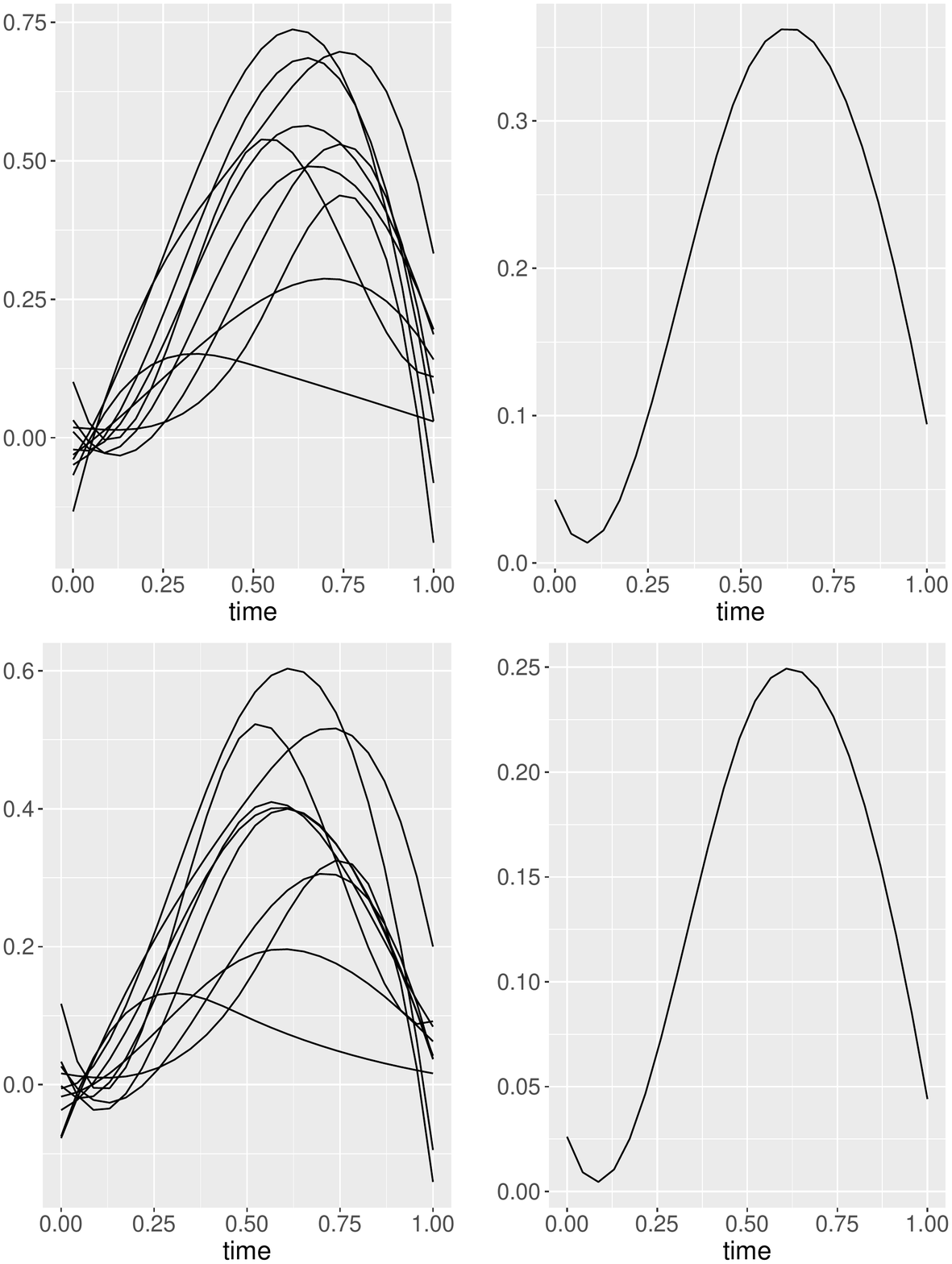}}
	\caption{Top row left panel contains the smooth function obtained from the raw physical activity intensity and the right panel depicts the mean of the smoothened data. Bottom left panel has the smooth functions obtained from the step count data. Bottom right panel is the mean of the step count functions.\label{smooth}}
\end{figure}

Recall that the application goal for the motivating data is to estimate the relationship between the true patterns of PA intensity $X(\cdot)$ and BMI. Since, the values of BMI are non-negative, we use the $log$ transformation i.e. consider the response $Y=log(BMI)$. The desired relationship is captured by $\tilde{\beta}(\cdot)$ in the regression model $Y=\tilde{\beta}_0+\int \tilde{\beta}(t)X(t)dt+e.$ The smooth function obtained from observed activity intensity is a proxy for the true intensity and is treated as $W(\cdot)$. The step count function is the instrumental variable. The error free covariates included in the analyses were age, sex, and race/ethnicity. To account for the covariates, we fit a linear model with $Y$ as the response and the covariates as the predictors.
We implement the naive and the proposed estimation method on the residuals from this linear model as the response and the functional data 
and conduct a test to check if the naive and corrected estimates are significant. The p-value for the proposed test using the test statistic $\hat{T}$ from Theorem \ref{thm3} is $0.01$. This indicates that the relationship is significant. The p-value for the naive test using $T_w$ in \eqref{naivetest} is $0.04.$ Thus, p-value for the corrected estimate is smaller than that for the naive one. Confidence intervals in Figure \ref{ci} indicate an inverse relationship between log BMI and true PA intensity level at the latter part of the day. An inverse relationship between log(BMI) and true PA intensity level provides some evidence that individuals who are more active with higher levels of PA intensity tend to have lower log(BMI) values. The timing of this finding can also potentially support the diurnal effects of PA, indicating that the timing of PA influences BMI values.



\begin{figure}[t]
	\centerline{\includegraphics[width=4in,height=3in]{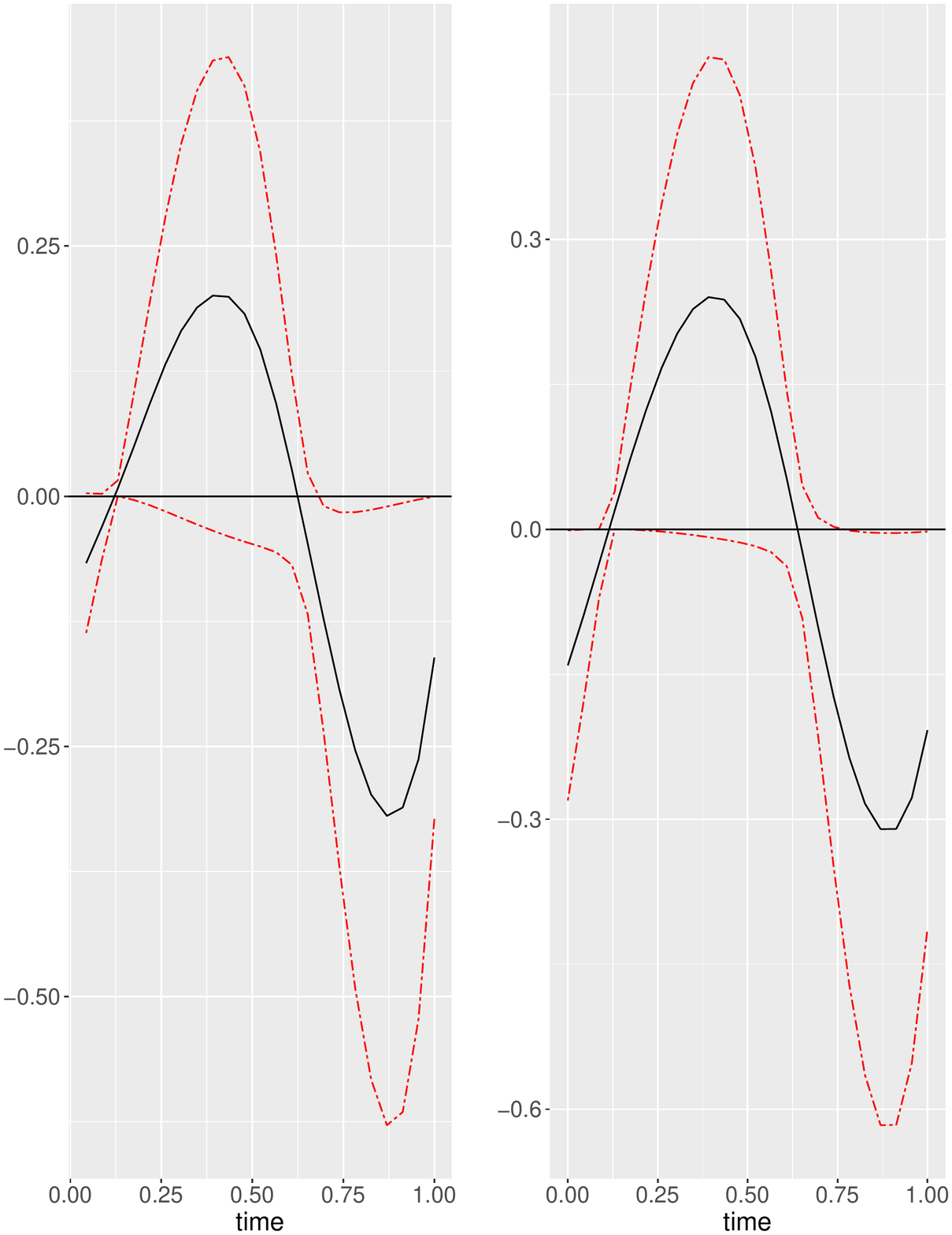}}
	\caption{Left figure has the corrected estimate and right one has the naive estimate. The confidence intervals are depicted in red.\label{ci}}
\end{figure}

   \section{Discussion}
With obesity becoming a growing public health concern, researchers are increasingly interested in determining how risk factors such as PA contributes to its development. To achieve this, PAM or wearable devices are employed to continuously monitor or track PA over a given time period. While it is well known that self-reported measures of PA are prone to errors, there remains additional work to be conducted in epidemiological studies to determine the accuracy device-based measures of PA in tracking 
PA. We successfully proposed a regression calibration-based method to correct for measurement error in the functional linear regression model by using extra available information on instrument variables.  As far as we know, the proposed method offers more general framework than existing methods to account for measurement errors using instruments. Based on our simulated data and theoretical results, we find that our proposed method performed better than current methods available for correcting for measurement error in this model setting.  Application of this method to the NHANES data set indicates that the relationship between patterns of PA intensity and BMI varies throughout the day. We generally observed an inverse relationship between PA and BMI towards the end of the day. Functional model offers a statistically sound framework to account for temporal factors that commonly used aggregate methods cannot. Though, the proposed method is quite generally applicable, it has potential for extensions. It can be extended to account for binary or more general responses i.e., develop a generalized linear functional model with measurement error. It can also be extended for data sets that are collected on family where the observations of family members are correlated i.e., multilevel models. 

\section{Data Availability Statement}
The data that support the findings of this study are available from the second author upon request.

\appendix

\section{Proof of Theorem 1}

We now prove the consistency of the estimate $\widehat{\widetilde{\theta}}_c$. For ease of writing, we use $c$ to denote a generic constant that can take different values on each occasion.

Recall that $\widehat{\widetilde{\theta}}_c=\left\{\sum_{i=1}^{n}\int Z_i^2(t)\phi(t)\phi^{'}(t)dt \right\}^{-1}\sum_{i=1}^{n} \int W_i(t)Z_i(t)\phi(t)dt.$ From (2) and (3) in the main paper, we obtain
$$\widehat{\widetilde{\theta}}_c= 
\left\{\sum_{i=1}^{n}\int Z_i^2(t)\phi(t)\phi^{'}(t)dt \right\}^{-1}\left[ \sum_{i=1}^{n} \int\left\{\theta(t)Z_i(t)+\widetilde{U}_i(t)+U_i(t) \right\}Z_i(t)\phi(t)dt\right].$$ Thus from the truncation, 
\begin{align}
\widehat{\widetilde{\theta}}_c-\widetilde{\theta}_c=\left\{\sum_{i=1}^{n}\int Z_i^2(t)\phi(t)\phi^{'}(t)dt \right\}^{-1}\left\{\int \sum_{i=1}^{n} U_{1i}(t)Z_i(t)\phi(t)dt\right\},
\end{align}

where, $U_{1i}(t)=\widetilde{U}_i(t)+U_i(t), i=1,...n.$ Let $A =\left\{n^{-1}\sum_{i=1}^{n}\int Z_i^2(t)\phi(t)\phi^{'}(t)dt \right\}^{-1}$ and
$b=(1/n)\left(\sum_{i=1}^{n}\int U_{1i}(t)Z_i(t)\phi_1(t)dt,...,\sum_{i=1}^{n}\int U_{1i}(t)Z_i(t)\phi_q(t)dt\right)'.$ Then, $\widehat{\widetilde{\theta}}_c-\widetilde{\theta}_c= Ab.$ 

we next prove some results necessary to prove Theorem 2.1.

\begin{lemma}\label{lem1}
	$ \|b\|=O_p(\sqrt{q/n})	$
\end{lemma}

\textbf{Proof}

We have 
\begin{align*}
\|b\|^2 &=n^{-2}\sum_{k=1}^{q} \left[\sum_{i=1}^{n} \int U_{1i}(t) Z_i(t)\phi_k(t)dt\right]^2\\
&=n^{-2} \sum_{k=1}^{q}\sum_{i=1}^{n} \left[\int U_{1i}(t) Z_i(t)\phi_k(t)dt\right]^2\\
&\quad +n^{-2}\sum_{k=1}^{q}\sum_{i_1=1}^{n}\sum_{i_2=1}^{n} \int U_{1i_i}(t) Z_{i_1}(t)\phi_k(t)dt\int U_{1i_2}(t) Z_{i_2}(t)\phi_k(t)dt\\
E(\|b\|^2)&=cq/n+0
\end{align*}

We use Assumption \ref{A1} to get the above result. Thus, the result is proved.

With the above Lemma, we are ready to prove Theorem 2.1.

\textbf{Proof}

\begin{align*}
\|\widehat{\widetilde{\theta}}_c-\widetilde{\theta}_c\|^2&=b'AAb\\
&=\lambda_{max} (AA) \|b\|^2\\
&\leq c\dfrac{q}{n}\\
\end{align*}
We use assumption \ref{A1}, and \ref{A3} and Lemma \ref{lem1} to get the above result. Thus, Theorem 1 is proved from Assumption \ref{A2}.

\section{Proof of Theorem 2}

We first prove some preliminary results.\\
Recall that $\widehat{\widetilde{\beta}}_c=(\sum_{i=1}^{n}\widehat{V}_{ci}\widehat{V}_{ci}^{'})^{-1}(\sum_{i=1}^{n}Y_i\widehat{V}_{ci}).$ From (1) in the main paper and the truncation, we obtain
$\widehat{\widetilde{\beta}}_c=(\sum_{i=1}^{n}\widehat{V}_{ci}\widehat{V}_{ci}^{'})^{-1}\left(\sum_{i=1}^{n}\widehat{V}_{ci}X_{ci}'\beta_c+\widehat{V}_{ci}e_i\right).$ Recall that $X_{ci}$ is the vector containing coefficients from the basis expansion of $ X_i(\cdot).$

\begin{lemma}\label{lem3}
	$\|\widehat{\widetilde{\beta}}_c-\widetilde{\beta}_c-(\sum_{i=1}^{n}\widehat{V}_{ci}\widehat{V}_{ci}^{'})^{-1}\left(\sum_{i=1}^{n}\widehat{V}_{ci}e_i\right)\|=o_p(1)$. Thus,\\ $\widehat{\widetilde{\beta}}_c-\widetilde{\beta}_c=(\sum_{i=1}^{n}\widehat{V}_{ci}\widehat{V}_{ci}^{'})^{-1}\left(\sum_{i=1}^{n}\widehat{V}_{ci}e_i\right)$ in probability.
\end{lemma}

\textbf{Proof}

\begin{align}
\widehat{\widetilde{\beta}}_c&=\left(\sum_{i=1}^{n}\widehat{V}_{ci}\widehat{V}_{ci}^{'}\right)^{-1}\left(\sum_{i=1}^{n}\widehat{V}_{ci}X_{ci}'\widetilde{\beta}_c+\widehat{V}_{ci}e_i\right) \nonumber\\
&=A+B, \label{eq1}
\end{align}
where $A$ and $B$ refer to the first and the second terms respectively after opening the bracket. Consider, the first term

\begin{align}
A&=\left(n^{-1}\sum_{i=1}^{n}\widehat{V}_{ci}\widehat{V}_{ci}^{'}\right)^{-1}n^{-1}\sum_{i=1}^{n}\widehat{V}_{ci}X_{ci}'\widetilde{\beta}_c \nonumber\\
&=\left(n^{-1}\sum_{i=1}^{n}\widehat{V}_{ci}\widehat{V}_{ci}^{'}\right)^{-1}n^{-1}\left[ \sum_{i=1}^{n}\widehat{V}_{ci}(X_{ci}'-\widehat{V}'_{ci}+\widehat{V}'_{ci})\widetilde{\beta}_c\right] \nonumber\\
&= \left(n^{-1}\sum_{i=1}^{n}\widehat{V}_{ci}\widehat{V}_{ci}^{'}\right)^{-1}n^{-1}\left[\sum_{i=1}^{n}\widehat{V}_{ci}\widehat{V}'_{ci}\beta_c+\widehat{V}_{ci}[X_{ci}'-\widehat{V}'_{ci}]\widetilde{\beta}_c\right] \nonumber\\
&=\widetilde{\beta}_c+\left(n^{-1}\sum_{i=1}^{n}\widehat{V}_{ci}\widehat{V}_{ci}^{'}\right)^{-1}\left(n^{-1}\sum_{i=1}^{n}\widehat{V}_{ci}[X_{ci}'-\widehat{V}'_{ci}]\widetilde{\beta}_c\right) \nonumber\\
&=\widetilde{\beta}_c+A_2.  \label{eq2}
\end{align}

Recall, $\widehat{V}_i(t)=\widehat{\widetilde{\theta}}(t)Z_i(t).$ Thus, from (3) in the main paper and the basis expansion the $j^{th}$ element of  $\widehat{V}'_{ci}$ is $\widehat{V}'_{cij}= \int \widehat{\widetilde{\theta}}(t)Z_i(t)\psi_j(t)dt,$ which gives $$ X_{cij}-\widehat{V}_{cij}=\int \{ \widetilde{\theta}(t)-\widehat{\widetilde{\theta}}(t)\}Z_i(t)\psi_j(t)dt+\int U_i(t)\psi_j(t)dt.$$

This yields, 

\begin{align*}
[X_{ci}-\widehat{V}_{ci}]^{'}\widetilde{\beta}_c&=\sum_{j=0}^{p} \widetilde{\beta}_j\int \{ \widetilde{\theta}(t)-\widehat{\widetilde{\theta}}(t)\}Z_i(t)\psi_j(t)dt\\
&\quad +\sum_{j=0}^{p}\widetilde{\beta}_j \int U_i(t)\psi_j(t)dt\\
n^{-1}\sum_{i=1}^{n}\widehat{V}_{ci}[X_{ci}'-\widehat{V}'_{ci}]\widetilde{\beta}_c&=(1/n)\sum_{i=1}^{n} \sum_{j=0}^{p}\widehat{V}_{ci} \widetilde{\beta}_j\int \{ \widetilde{\theta}(t)-\widehat{\widetilde{\theta}}(t)\}Z_i(t)\psi_j(t)dt\\
&\quad +(1/n)\sum_{i=1}^{n} \sum_{j=0}^{p} \widehat{V}_{ci}\widetilde{\beta}_j \int U_i(t)\psi_j(t)dt\\
&=T_1 +T_2
\end{align*} 

We will shortly show that $\|T_1\|, \|T_2\| =o_p(1).$ This result yields $  A_2=\left(n^{-1}\sum_{i=1}^{n}\widehat{V}_{ci}\widehat{V}_{ci}^{'}\right)^{-1}(T_1+T_2)$. So, $\|A_2\|^2\leq \lambda_{max}\left[\left(n^{-1}\sum_{i=1}^{n}\widehat{V}_{ci}\widehat{V}_{ci}^{'}\right)^{-2}\right](\|T_1+T_2\|^2)=o_p(1),$ from Assumption \ref{A7}. This, along with \eqref{eq1}, \eqref{eq2} proves the Lemma.

We now show that $\|T_1\|, \|T_2\| =o_p(1).$

\begin{align*}
T_1&=(1/n)\sum_{i=1}^{n} \sum_{j=0}^{p}\widehat{V}_{ci} \widetilde{\beta}_j\int \{ \widetilde{\theta}(t)-\widehat{\widetilde{\theta}}(t)\}Z_i(t)\psi_j(t)dt\\
&=(1/n)\sum_{i=1}^{n} \widehat{V}_{ci} \int \{ \widetilde{\theta}(t)-\widehat{\widetilde{\theta}}(t)\}Z_i(t)\sum_{j=0}^{p}\widetilde{\beta}_j\psi_j(t)dt\\
&\leq (1/n)\sum_{i=1}^{n} \|\widehat{V}_{ci}\| \| \int\{ \widetilde{\theta}(t)-\widehat{\widetilde{\theta}}(t)\}Z_i(t)dt\|\\
\|T_1\|&\leq c  \dfrac{\sqrt{q}}{\sqrt{n}}
\end{align*}

We use Assumption \ref{A2} and Theorem 2.1 to get the above results.
Next, consider

\begin{align*}
T_2&=(1/n)\sum_{i=1}^{n} \widehat{V}_{ci}\int  U_i(t)\sum_{j=0}^{p} \widetilde{\beta}_j\psi_j(t)dt=(1/n)\sum_{i=1}^{n} \widehat{V}_{ci}\int  U_i(t)B(t)dt,
\end{align*}

where $\sum_{j=0}^{p} \widetilde{\beta}_j\psi_j(t)=B(t).$ Note, that $\|B(\cdot)\|_L <c.$ Let $ \int U_i(t)B(t)dt=\mathcal{U}_i.$ Note that $\mathcal{U}_i$ are i.i.d., centered and independent of $V_{ic}$ with $E\|\mathcal{U}_i\|^2 <\infty.$

\begin{align*}
T_2 &=(1/n)\sum_{i=1}^{n} \widehat{V}_{ci}\mathcal{U}_i\\
&=(1/n)\sum_{i=1}^{n} (\widehat{V}_{ci}-X_{ci})\mathcal{U}_i+(1/n)\sum_{i=1}^{n} X_{ci}\mathcal{U}_i\\
E(\|T_2\|^2)&=cq/n
\end{align*}

Thus, $\|T_1\|, \|T_2\| =o_p(1).$

\textbf{Proof of Theorem 2}

We now have that $\widehat{\widetilde{\beta}}_c-\widetilde{\beta}_c=(\sum_{i=1}^{n}\widehat{V}_{ci}\widehat{V}_{ci}^{'})^{-1}\left(\sum_{i=1}^{n}\widehat{V}_{ci}e_i\right)$. Let $D=(\sum_{i=1}^{n}\widehat{V}_{ci}\widehat{V}_{ci}^{'})^{-1}$ and $b=\sum_{i=1}^{n}\widehat{V}_{ci}e_i$. Similar to Lemma \ref{lem1}, we can show that $\|b\|=O_p(\sqrt{p/n}). $

\begin{align*}
\|\widehat{\widetilde{\beta}}_c-\widetilde{\beta}_c\|^2&=b'DDb\\
&\leq \lambda_{max} (DD) \|b\|^2\\
&\leq c\dfrac{p}{n}\\
\end{align*}
We use Assumptions \ref{A5} and \ref{A7} and get the above result prove Theorem 2.



\bibliographystyle{abbrv}

\bibliography{manu_1}

\begin{thebibliography}{10}

\bibitem{benatia2017functional}
D.~Benatia, M.~Carrasco, and J.-P. Florens.
\newblock Functional linear regression with functional response.
\newblock {\em Journal of econometrics}, 201(2):269--291, 2017.

\bibitem{cai2015methods}
X.~Cai.
\newblock {\em Methods for handling measurement error and sources of variation
  in functional data models}.
\newblock PhD thesis, Columbia University, 2015.

\bibitem{cardot2007smoothing}
H.~Cardot, C.~Crambes, A.~Kneip, and P.~Sarda.
\newblock Smoothing splines estimators in functional linear regression with
  errors-in-variables.
\newblock {\em Computational statistics \& data analysis}, 51(10):4832--4848,
  2007.

\bibitem{carroll2006measurement}
R.~J. Carroll, D.~Ruppert, L.~A. Stefanski, and C.~M. Crainiceanu.
\newblock {\em Measurement error in nonlinear models: a modern perspective}.
\newblock CRC press, 2006.

\bibitem{chakraborty2017regression}
A.~Chakraborty and V.~M. Panaretos.
\newblock Regression with genuinely functional errors-in-covariates.
\newblock {\em arXiv preprint arXiv:1712.04290}, 2017.

\bibitem{chiou2004functional}
J.-M. Chiou, H.-G. M{\"u}ller, and J.-L. Wang.
\newblock Functional response models.
\newblock {\em Statistica Sinica}, pages 675--693, 2004.

\bibitem{chung2012physical}
A.~E. Chung, A.~C. Skinner, M.~J. Steiner, and E.~M. Perrin.
\newblock Physical activity and bmi in a nationally representative sample of
  children and adolescents.
\newblock {\em Clinical pediatrics}, 51(2):122--129, 2012.

\bibitem{church2018obesity}
T.~Church and C.~K. Martin.
\newblock The obesity epidemic: a consequence of reduced energy expenditure and
  the uncoupling of energy intake?
\newblock {\em Obesity}, 26(1):14--16, 2018.

\bibitem{corder2007accelerometers}
K.~Corder, S.~Brage, and U.~Ekelund.
\newblock Accelerometers and pedometers: methodology and clinical application.
\newblock {\em Current Opinion in Clinical Nutrition \& Metabolic Care},
  10(5):597--603, 2007.

\bibitem{goldsmith2011penalized}
J.~Goldsmith, J.~Bobb, C.~M. Crainiceanu, B.~Caffo, and D.~Reich.
\newblock Penalized functional regression.
\newblock {\em Journal of computational and graphical statistics},
  20(4):830--851, 2011.

\bibitem{jadhav2017dependent}
S.~Jadhav, H.~Koul, and Q.~Lu.
\newblock Dependent generalized functional linear models.
\newblock {\em Biometrika}, 104(4):987--994, 2017.

\bibitem{jadhav2020functional}
S.~Jadhav and S.~Ma.
\newblock Functional measurement error in functional regression.
\newblock {\em Canadian Journal of Statistics}, 48(2):238--258, 2020.

\bibitem{jago2005bmi}
R.~Jago, T.~Baranowski, J.~C. Baranowski, D.~Thompson, and K.~Greaves.
\newblock Bmi from 3--6 y of age is predicted by tv viewing and physical
  activity, not diet.
\newblock {\em International journal of obesity}, 29(6):557--564, 2005.

\bibitem{john2012actigraph}
D.~John and P.~Freedson.
\newblock Actigraph and actical physical activity monitors: a peek under the
  hood.
\newblock {\em Medicine and science in sports and exercise}, 44(1 Suppl 1):S86,
  2012.

\bibitem{johnson2013national}
C.~L. Johnson, R.~Paulose-Ram, C.~L. Ogden, M.~D. Carroll, D.~Kruszan-Moran,
  S.~M. Dohrmann, and L.~R. Curtin.
\newblock National health and nutrition examination survey. analytic
  guidelines, 1999-2010.
\newblock 2013.

\bibitem{kim2018additive}
J.~S. Kim, A.~Maity, and A.-M. Staicu.
\newblock Additive nonlinear functional concurrent model.
\newblock {\em Statistics and its interface}, 11(4):669, 2018.

\bibitem{kokoszka2017introduction}
P.~Kokoszka and M.~Reimherr.
\newblock {\em Introduction to functional data analysis}.
\newblock CRC press, 2017.

\bibitem{kozey2010comparison}
S.~L. Kozey, J.~W. Staudenmayer, R.~P. Troiano, and P.~S. Freedson.
\newblock A comparison of the actigraph 7164 and the actigraph gt1m during
  self-paced locomotion.
\newblock {\em Medicine and science in sports and exercise}, 42(5):971, 2010.

\bibitem{leroux2019organizing}
A.~Leroux, J.~Di, E.~Smirnova, E.~J. Mcguffey, Q.~Cao, E.~Bayatmokhtari,
  L.~Tabacu, V.~Zipunnikov, J.~K. Urbanek, and C.~Crainiceanu.
\newblock Organizing and analyzing the activity data in nhanes.
\newblock {\em Statistics in biosciences}, 11(2):262--287, 2019.

\bibitem{ma2016dynamic}
H.~Ma, Y.~Bai, and Z.~Zhu.
\newblock Dynamic single-index model for functional data.
\newblock {\em Science China Mathematics}, 59(12):2561--2584, 2016.

\bibitem{maity2017nonparametric}
A.~Maity.
\newblock Nonparametric functional concurrent regression models.
\newblock {\em Wiley Interdisciplinary Reviews: Computational Statistics},
  9(2):e1394, 2017.

\bibitem{muller2005generalized}
H.-G. M{\"u}ller, U.~Stadtm{\"u}ller, et~al.
\newblock Generalized functional linear models.
\newblock {\em Annals of Statistics}, 33(2):774--805, 2005.

\bibitem{ng2014global}
M.~Ng, T.~Fleming, M.~Robinson, B.~Thomson, N.~Graetz, C.~Margono, E.~C.
  Mullany, S.~Biryukov, C.~Abbafati, S.~F. Abera, et~al.
\newblock Global, regional, and national prevalence of overweight and obesity
  in children and adults during 1980--2013: a systematic analysis for the
  global burden of disease study 2013.
\newblock {\em The lancet}, 384(9945):766--781, 2014.

\bibitem{ramsay2004functional}
J.~O. Ramsay.
\newblock Functional data analysis.
\newblock {\em Encyclopedia of Statistical Sciences}, 4, 2004.

\bibitem{reiss2017methods}
P.~T. Reiss, J.~Goldsmith, H.~L. Shang, and R.~T. Ogden.
\newblock Methods for scalar-on-function regression.
\newblock {\em International Statistical Review}, 85(2):228--249, 2017.

\bibitem{robertson2011utility}
W.~Robertson, S.~Stewart-Brown, E.~Wilcock, M.~Oldfield, and M.~Thorogood.
\newblock Utility of accelerometers to measure physical activity in children
  attending an obesity treatment intervention.
\newblock {\em Journal of obesity}, 2011, 2011.

\bibitem{sera2017using}
F.~Sera, L.~J. Griffiths, C.~Dezateux, M.~Geraci, and M.~Cortina-Borja.
\newblock Using functional data analysis to understand daily activity levels
  and patterns in primary school-aged children: cross-sectional analysis of a
  uk-wide study.
\newblock {\em PLoS one}, 12(11):e0187677, 2017.

\bibitem{tekwe2019instrumental}
C.~D. Tekwe, R.~S. Zoh, M.~Yang, R.~J. Carroll, G.~Honvoh, D.~B. Allison,
  M.~Benden, and L.~Xue.
\newblock Instrumental variable approach to estimating the scalar-on-function
  regression model with measurement error with application to energy
  expenditure assessment in childhood obesity.
\newblock {\em Statistics in medicine}, 38(20):3764--3781, 2019.

\bibitem{xu2019modeling}
S.~Y. Xu, S.~Nelson, J.~Kerr, S.~Godbole, E.~Johnson, R.~E. Patterson, C.~L.
  Rock, D.~D. Sears, I.~Abramson, and L.~Natarajan.
\newblock Modeling temporal variation in physical activity using functional
  principal components analysis.
\newblock {\em Statistics in Biosciences}, 11(2):403--421, 2019.

\bibitem{yao2005functional}
F.~Yao, H.-G. M{\"u}ller, and J.-L. Wang.
\newblock Functional data analysis for sparse longitudinal data.
\newblock {\em Journal of the American statistical association},
  100(470):577--590, 2005.

\end{thebibliography}

\end{document}